\documentclass[%
 reprint,
superscriptaddress,
 amsmath,amssymb,
 aps,
]{revtex4-2}

\usepackage{graphicx}
\usepackage{dcolumn}
\usepackage{bm}
\usepackage{cases}
\usepackage{xcolor}
\usepackage{soul}
\usepackage{amsmath,amsfonts}

\begin{document}

\preprint{APS/123-QED}

\title{General model for charge carriers transport in electrolyte-gated transistors}

\author{Marcos Luginieski}
 \affiliation{Instituto de Física de São Carlos, Universidade de São Paulo, CP 369, CEP 13660-970, São Carlos, SP, Brazil}
  \affiliation{Universidade Tecnológica Federal do Paraná - UTFPR, Av. Sete de Setembro, 3165, CEP 80230-901 Curitiba, Brazil}
 \email{mluginieski@gmail.com }
\author{Marlus Koehler}%
 \affiliation{Universidade Federal do Paraná - UFPR, Centro Politécnico, Jardim das Américas CP 19044, CEP 81531-990 Curitiba,
Brazil}
%
\author{José P. M. Serbena}
 \affiliation{Universidade Federal do Paraná - UFPR, Centro Politécnico, Jardim das Américas CP 19044, CEP 81531-990 Curitiba,
Brazil}%
\author{Keli F. Seidel}
\email{keliseidel@utfpr.edu.br}
 \affiliation{Universidade Tecnológica Federal do Paraná - UTFPR, Av. Sete de Setembro, 3165, CEP 80230-901 Curitiba, Brazil}%

\date{\today}

\begin{abstract}
Inspired by experimental observations related to electrolyte-gated transistors (EGTs) where non-ideals behaviors are shown and not described by just one theoretical model, we proposed a charge carriers transport model able to describe the typical modes of operation profiles as well as some non-ideals ones from electrolyte-gated field effect transistors (EGOFETs) and organic electrochemical transistors (OECTs). Our analysis include the effect of 2D or 3D percolation transport (PT) and also the influence of a shallow exponential traps distribution on the transport. Under these considerations, a non-constant accumulation layer thickness along the channel can be formed. Such dependence was included into our model in the effective mobility parameter dependent on the accumulation thickness. The accumulation thickness can depict 2D or 3D PT or even a transition between them. This transition can produce a non-ideal profile between the linear and saturation regimes in the output curve, region in which a protuberance/lump appears. Other analyzed phenomenon was the non-linear behavior for low drain voltage range in the output curve, even when considering an ohmic contact. According to this proposed model, this curve behavior is attributed to the traps distribution profile into the semiconductor and the very thin accumulation layer thickness close to the injection contact. It was also possible to analyze the conditions when the linear field effect mobility ($\mu_{\textrm{lin}}$) is higher or lower than the saturation one ($\mu_{\textrm{sat}}$).
Finally, EGOFET and OECT experimental data were successfully fitted with this model showing its versatility. 
\end{abstract}

\maketitle


\section{\label{introduction}Introduction}
\par There are many well established phenomenological models to describe the charge carrier transport in field effect transistors (FETs) \cite{vanhutten2000}. When such models are extended to organic field effect transistors (OFETs) or even to electrolyte-gated transistors (EGTs), new kinds of theoretical schemes were proposed by adapting models originally developed for silicon-based FETs \cite{inbook_horowitz2010, ART_delavari2021, ART_bernards2007, bao2007}. 
General phenomenological models that encompass the peculiarities of OFETs or EGTs are more uncommon in the literature. Specifically about EGTs, their theoretical description is more complex since those devices
have two distinguished typical modes of operation: (i) transconductance due to field effect only being named as electrolyte-gated organic field effect transistors (EGOFETs) or due to (ii) ionic current being named as organic electrochemical transistors (OECTs).  In spite of that, it is a common practice in the literature to extract the EGOFET parameters by using approaches based on the OFET models  \cite{inbook_horowitz2010}. 
\par Furthermore, there are theoretical approaches specifically created for EGOFETs, e.g, a model based on the Nernst-Plank-Poisson equation \cite{ART_delavari2021}. This model spatially maps the concentration of electrons and ions, together with the distribution of electric potential and field along the transistor channel. Yet it does not provide an analytical equation for the drain current that can be straightforwardly compared to experiments. It also has the additional disadvantage to be based on non-free software. For OECTs, one of the main description was proposed in 2007 by Bernards and Malliaras \cite{ART_bernards2007}, where the doping or de-doping level is expressed through a volumetric analysis of the organic semiconductor modeled by a gate capacitance ($C_G$). Some models also consider a temporal variation of the ionic current into the channel and its transient from the doped to de-doped (or vice-versa) promoted by the diffused ions from the electrolyte to the channel \cite{ART_bernards2007,ART_colucci2020}. All these proposed theories are specifically applied to EGOFETs or OECTs, but they are not universal approaches since the extraction of characteristic parameter from electrolyte-gated transistors still depends on {\it a priori} assumption of the operation mode.
\par
\begin{figure}
        \centering
        \includegraphics[width=0.9\linewidth]{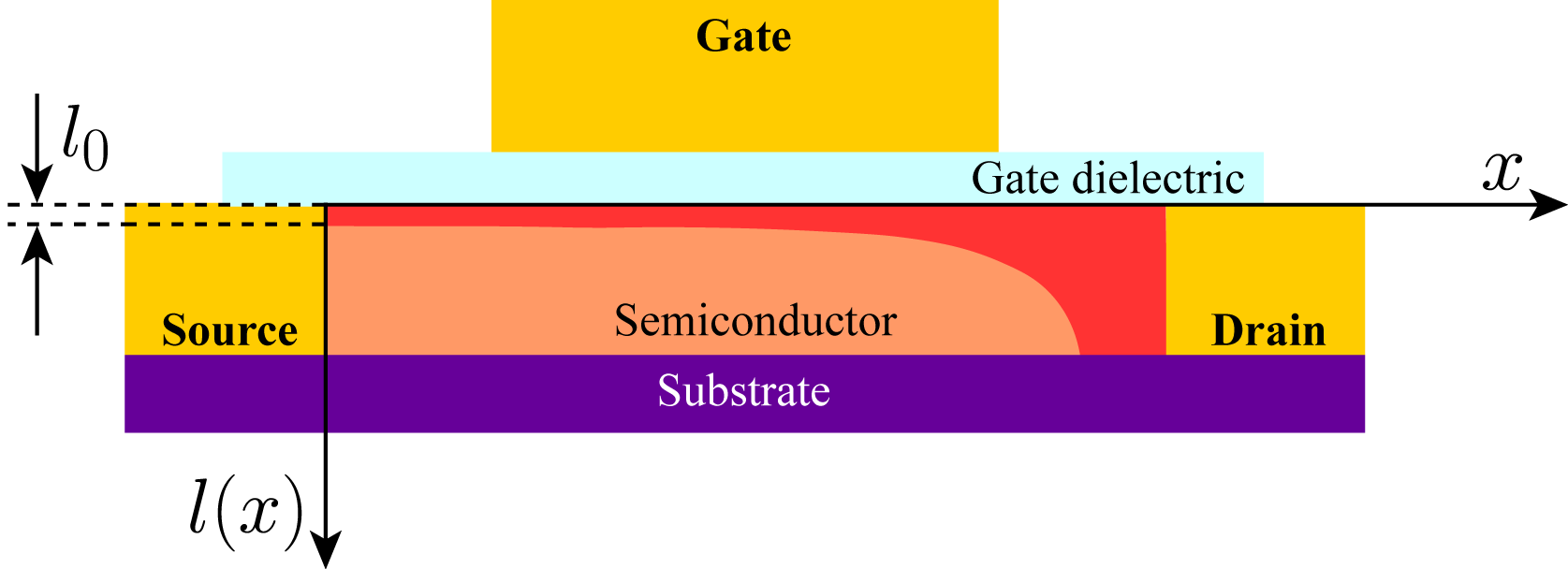}
        \caption{Field effect transistor architecture with bottom contact and top gate and its accumulation thickness variation along the semiconductor channel length (dark red highlights) into the channel. $l_0$ is the minimum effective thickness.}
        \label{fig:variacao_l}
    \end{figure}

 Here we propose that all those gaps can be filled by considering a model for an injection-based surface field effect transistors (IFETs) \cite{ART_koehler2010}. The architecture of the IFET is comparable to thin film transistors (TFTs) or OFETs as shown in figure \ref{fig:variacao_l}. The model analyzes the charge transport by considering the competition between charge carrier trapping at an exponential distribution of traps and the transport improvement produced by the formation of 2D or 3D percolation pathways along the channel. The influence of these two effects was introduced on the effective mobility that is equivalent to the field effect mobility, given by \cite{ART_koehler2010}:
\begin{equation}
        \mu_{\textrm{eff}} \propto \left(D - D_c\right)^{\alpha} D^{-(\gamma - 1)}, \quad D \geq D_{c}.
        \label{eq_mobility_adimentional}
\end{equation}
where $\alpha$ is a parameter correlated to the dimension of the  percolation pathways, $D_c$ a minimum thickness of the accumulation layer necessary to establish a percolation pathway between source and drain, and $D$ the thickness of the semiconductor film. Necessarily, Eq. \eqref{eq_mobility_adimentional} is valid only when $D> D_c$. $\alpha$ can assume values between $1-1.4$ for 2D percolation transport (PT) and $1.5-2$ for 3D PT \cite{ART_dinelli2004,ART_koehler2010}. The parameter $\gamma$ in Eq. \eqref{eq_mobility_adimentional} represents the energetic depth of the exponential distribution of traps. It is usually written as $\gamma = T_c/T$, where $T_c$ is the characteristic temperature of the trap distribution, and $T$ is the temperature. Equation \ref{eq_mobility_adimentional} was
written by assuming the mobility in a percolative
problem with $D$ following then a power law of the kind $\mu_\textrm{eff} \propto (D - Dc)^{\alpha}$ \cite{ART_koehler2010}.
On the other hand,  the mobility in the conducting islands decreases with increasing film thickness following also a power law of the kind $\mu_\textrm{eff} \propto D^{-(\gamma - 1)}$. This dependence is due to the trap-limited transport within the island, so that the mobility decreases with the increasing
number of traps in thicker layers. Hence, the mobility rise due to improved island contacts with increasing $D$  can be partially compensated by the poor transport in those islands.
\par Another phenomenon known in FETs and EGTs literature is that the accumulation layer does not necessarily occupies the entire thickness of the semiconductor film \cite{ART_koehler2010,OECT_channel_thickness_exponential}, which can be particularly important in EGTs because these devices have  a high capacitance per unit area produced by the electrolyte dielectric film. The density of induced charges on the semiconductor/dielectric surface is then high even at low applied voltages. Due to this key feature, a proper modelling of the charge transport in EGT must take into account the effects produced by the variations in the thickness of the accumulation layer ($l$) with the gate voltage. Since $l$ might abruptly change from just a few nanometers up to the entire thickness of the semiconductor film thickness with the voltage variation, it is expected that the gate voltage can induce transitions on the regime of transport. The percolation pathways between source and drain can then present a transition from 2D to 3D transport for a determined voltage, for instance. Such transition
might also have a complex behaviour produced by the competition between this effect (gauged by the parameter $\alpha$ in Eq. \eqref{eq_mobility_adimentional}) and the trap filling process that depends on $\gamma$.
\par In this way, the present work expands and extends the analysis initially developed in Ref. \cite{ART_koehler2010} to derive a single model for EGTs. From that model, it is possible to obtain valuable information on typical parameters that controls the behaviour of those transistors, using a unique theoretical framework to study the EGOFETs and OECTs modes of operation. In addition to the typical macroscopic parameters analyzed, from the present model it is also possible to infer: (i) the profile of the accumulation layer thickness that does not necessarily extend along all the semiconductor film; (ii) when the linear mobility is higher or lower than the saturation one (or vice-versa); (iii) the non-constant field effect mobility dependent on accumulation layer thickness ($\mu_{\textrm{eff}}(l)$); (iv) the non-linear regime in the output curve even when there is an ohmic-contact; (v) peculiar profiles in the output curve attributed to the transition from 2D to 3D percolation transport.
\section{\label{model}Theoretical model}
\par  The model is reasoned for a transistor architecture as depicted in figure \ref{fig:variacao_l} with bottom contacts and top gate. We are mainly interested in properly describe the charge transport effects associated with the variations on the thickness of the accumulation layer ($l$) along the channel. Ref. \cite{ART_koehler2010} initially addressed the problem to calculate $l$ by solving the drift-diffusion and 1D Poisson equations along the $y$-direction (the direction is perpendicular to the semiconductor/dielectric interface, see Fig. \ref{fig:variacao_l} ). They assumed equilibrium condition under
the gradual channel and trap-free approximations. Since the component of the electric-field in the $y$-direction is much more intense than the component in $x$-direction, the electric potential ($V$) is then a function of the direction parallel to semiconductor/dielectric interface ($x$-direction) \cite{ART_koehler2010,gradual_channel}. Those assumptions implies that the current flow occurs only along the plane parallel to the electrolyte/semiconductor interface. As a result, the effective thickness of the accumulation layer ($l(x)$) is a function of $V(x)$ at a coordinate $x$ in the channel: 
\begin{equation}
     l (x) = \begin{cases}
           \left( \frac{V_\textrm{tr}}{V_{G} -V_T - V(x)}\right)\,D, \quad &V(x) < V_{\textrm{sat}} - V_\textrm{tr},
            \\
            \\
           \,\, D, \quad &V(x) \geq V_{\textrm{sat}} - V_\textrm{tr},
         \label{eq_effective_thickeness}
        \end{cases}
\end{equation}
  where $V_{\textrm{sat}} \equiv V_G-V_T$, $V_T$ is the threshold voltage, $V_G$ the gate voltage, $V_\textrm{tr} \equiv 4\epsilon k T_c/eC_i D$, $\epsilon$ is the electrical permittivity of the semiconductor and $C_i$ is the capacitance per unit area of the gate dielectric.  The variation of $l(x)$ as a function of $V(x)$ is schematically illustrated in figure \ref{fig:variacao_l} where the dark red profile filling the channel represents the accumulation layer. At the source region, $V$(0)\,=0$\,$V, the thickness of the accumulation layer has its minimum value (called minimum effective thickness), given by
    $l_{0} = D\{ V_\textrm{tr}/V_{\textrm{sat}}\}$.
    \par In Ref. \cite{ART_koehler2010} the interplay between the percolation and the trap-filling process to determine the  effective mobility ($\mu_{\textrm{eff}}$) of an IFET was written as a function of the entire semiconductor thickness ($D$). In an EGTs, however, it is expected that the accumulation layer does not necessarily extend along the entire $D$, but remains concentrated near the semiconductor/dielectric interface for a wide range of voltage operation.  It is then reasonable to introduce an effective mobility that depends on the thickness of the accumulation layer by writing $\mu_{\textrm{eff}}$ as a function of $l$ instead of $D$ in Eq. \eqref{eq_mobility_adimentional}, where $l$ is given by Eq. \eqref{eq_effective_thickeness}: 
    \begin{equation}
        \mu_{\textrm{eff}}(l_x) = \mu_\textrm{sat}\left[\frac{(l - D_{c})}{D}\right]^{\alpha} \left(\frac{l}{D}\right)^{-(\gamma - 1)},\, D_c < l \leq D.
        \label{Eq:mob2}
    \end{equation}
    where $ \mu_\textrm{sat}$ is the field effect mobility in the saturation regime given by a constant so that $\mu_{\textrm{eff}} = \mu_\textrm{sat}$ when $l = D \gg D_c$. The above equation is defined for $l>D_c$ since the charge transport between source and drain is only possible if there is at least one percolation path linking both electrodes. 
    \par From the conditions in Eq. \eqref{eq_effective_thickeness}, the  $\mu_{\textrm{eff}}$ determined by Eq. \eqref{Eq:mob2} is valid when  $V(x) < V_{\textrm{sat}} - V_{\textrm{tr}}$. On the other hand, when $V(x) > V_{\textrm{sat}} - V_{\textrm{tr}}$, the effective mobility has a constant value
    \begin{equation}
        \mu_{\textrm{eff}}(l = D) = \mu_{\textrm{sat}}\left(\frac{D - D_c}{D}\right)^\alpha.
        \label{Eq:mob3}
    \end{equation}
    \par The drain current for a device illustrated in figure \ref{fig:variacao_l} is calculated from the density of free charge carriers induced in the channel by $V_G$, or $|dq|=|C_i\,W(V_{\textrm{sat}}-V(x))dx|$ \cite{inbook_horowitz2010}, where $W$ is the channel width. One can also write  $I_D=dq/dt=(dx/dt)\cdot (dq/dx)=\mu_{\textrm{\textrm{eff}}}(dV/dx)(dq/dx)$, where  $\mu_{\textrm{eff}}$ is the ratio between the charge carriers' velocity and the component of the electric field along the $x$-direction. The expression of the drain current is then \cite{inbook_horowitz2010, ART_koehler2010}:
     \begin{equation}
        I_{D} = \frac{W C_i}{L}\int_{0}^{V_{D}} \mu_{\textrm{eff}}(l) [V_G - V_{T} - V(x)] dV.
        \label{eq_Ids_base}
    \end{equation}
    \par Eq. \eqref{eq_Ids_base} can be calculated as a function of $V_D$ or $V_G$ to give the device's output or transfer typical curves, respectively. Since the field effect mobility is not constant in this model, the integration of Eq. \eqref{eq_Ids_base} must be performed taking into account the two limits settled by Eq. \eqref{Eq:mob2}
    \par For simplicity (without loss of generality), we are going to assume that the device in figure \ref{fig:variacao_l} works only when positive bias is applied to the source/gate electrodes (it means, an n-type EGT). The  \textbf{linear regime} of operation is established when $V_D < V_{\textrm{sat}}$. The integration of Eq. \eqref{eq_Ids_base} for this regime gives:
    \begin{equation}
        I_{D} = \frac{\mu_\textrm{sat}C_{i}W}{LD^{\alpha}}\left[V_{\textrm{tr}}^{-(\gamma-1)}\Lambda+\\\left(D - D_c\right)^{\alpha}\Phi\right],\\ \label{id}
    \end{equation}
    where the $\Lambda$ in Eq. \eqref{id} is  defined as    %
    \begin{equation}
        \Lambda(V^*) =D^{\alpha}\int_{0}^{V^*}\left[\frac{V_{\textrm{tr}}}{V_{\textrm{sat}} - V(x)} - \frac{D_c}{D}\right]^{\alpha} [V_{\textrm{sat}} - V(x)]^{\gamma} dV.\label{lambda}
    \end{equation}
    \par The integral in Eq. \eqref{lambda} cannot be solved analytically, so it needs to be solved numerically. The limit of integration ($V^*$) in this equation depends on the voltage $V''=V_{\textrm{sat}} - V_{\textrm{tr}}$: if $V_D < V''$, $V^*=V_D$ otherwise $V^*=V''$. Finally, the $\Phi$ in Eq. \eqref{id} is a function of $V_D$ and $V''$, that is,
    \begin{equation}
        \begin{split}
            \Phi(V'',V_D) =\theta(V_D-V'') \left[V_{\textrm{sat}}(V_{D} - V'') - \frac{V_{D}^{2} - V''^{2}}{2}\right],
        \end{split}
    \end{equation}
    where $\theta(V_D-V'')$ is the Heaviside's step function.
    \par The \textbf{saturation regime} occurs when $V_{D} \geq V_{\textrm{sat}}$. Under this regime the drain current is:
    \begin{equation}
        I_{D,\textrm{sat}} = I_D(V'',V_{\textrm{sat}}) + \frac{\mu_\textrm{sat}C_{i}W}{L}  \left(\frac{D - D_c}{D}\right)^{\alpha} \frac{V_\textrm{tr}^2}{2}, 
    \end{equation}
    where $I_D(V'',V_{\textrm{sat}})$ is the current given by Eq. \eqref{id} with $V_D=V_{\textrm{sat}}$.
\section{Results}
\label{results}
    \par
\begin{figure}
        \centering
        \includegraphics[width=\columnwidth]{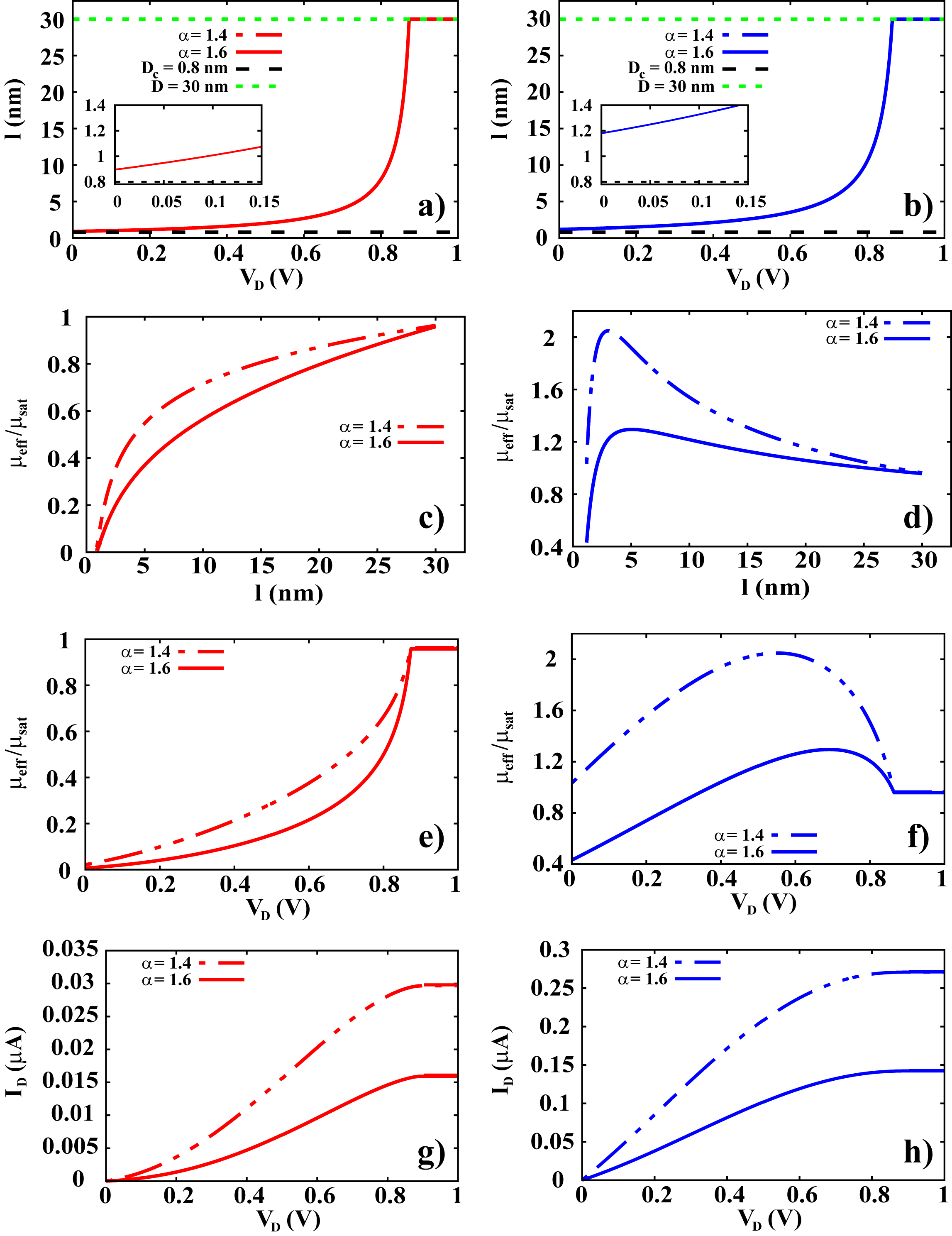}
        \caption{Electrolyte-gated transistor behaviors for 2D percolation transport (2D PT - dashed lines) and 3D percolation transport (3D PT - solid lines). a) - b) Variation of the accumulation layer thickness $l=l_x$ as a function of $V_D$. c) - d) and e) - f) Effective mobility normalized to the saturation one ($\mu_\textrm{eff}/\mu_\textrm{sat}$) as a function of $l$ and $V_D$, respectively. g) - h) Output curve ($I_D - V_D$). The transistors parameters used for this  simulations are: $\alpha = 1.4$ for 2D PT and $\alpha = {1.6}$ for 3D PT, $D={30}\,\rm{nm}$,  $D_c=0.8\, \rm{nm}$, $W=1.0\, \rm{mm}$, $L = 30 \rm{\mu m}$, $\kappa = 4.0$, $T = 300\,\rm{ K}$, $V_G = 1.0\,\rm{ V}$, $V_T = 0.1\,\rm{ V}$ and $\mu_{\textrm{sat}} = 1.23\times 10^{-2}\rm{cm^2/Vs}$.}
        \label{fig:EGOFET_2D}
    \end{figure}

 Through the proposed model, we effortlessly analyzed some phenomena not easily obtained experimentally. Figure \ref{fig:EGOFET_2D} summarises the first results, where used parameters values are close to the observed in EGOFETs     \cite{ART_kergoat2012_PNAS_non_linear_fit, water_gated_transistor_horowitz_2010, Seidel_2021_transistor_Elton}. It was compared the transistor behaviour under 2D percolation transport (PT) regime (dashed lines) and 3D one (solid lines). The 2D PT can occur when the electronic coupling between states belonging to the same layer is higher compared to the coupling between states belonging to adjacent layers of the semiconductor film. In this situation, the in-plane percolation paths are formed with higher probability. The charge carriers tend then to flow along the same plane of the semiconductor film. For the 3D PT, however, the in-plane and inter-plane percolation paths can occur with the same probability since the electronic coupling between states of the same layer is comparable to the coupling between states of adjacent layers.  Here, the main parameter to make a distinction between an OFET and an EGOFET is the capacitance per unit area ($C_i$). The chosen value used was $C_i=1.0\,\rm{ \mu F/cm^2}$, since values around $1.0\,\rm{ \mu F/cm^2}$ to $10\,\rm{ \mu F/cm^2}$ are typically measured for EGOFETs. 
    \par First, the graphs from the left column side in Figure \ref{fig:EGOFET_2D} will be discussed, considering $\gamma=2.2$ represented by red curves. Figure \ref{fig:EGOFET_2D} a) shows the accumulation thickness \textit{versus} the applied drain voltage ($l - V_D$). The green dashed curve indicates a semiconductor film thickness $D$ and the black dashed one shows the critical percolation thickness $D_c$. 
    At zero volts is the grounded source electrode ($V(x=0)= 0$ V) and at $V(x=L$)=$V_D$= 1V is the drain electrode position. 
    The result shows that $l$ is non-constant along the channel length and does not extend along all the semiconductor thickness until $\sim 0.85\,V$, under the simulated parameters. Close to the source electrode, the accumulation layer has just a few nanometers (see inset) and it reaches the thickness $D$ close to the drain. If the same simulation is repeated just changing the $C_i$ value to one similar to that observed in OFETs ($\sim nF/cm^2$), $l = D$ thickness for all the channel length (see supplementary material). It proves that the analysis of the accumulation layer thickness is quite relevant for electrolyte-gated transistors due to its high $C_i$. 
    \par The graphs c) and e) in Figure \ref{fig:EGOFET_2D} present the normalized ratio of $\mu_\textrm{eff}/\mu_\textrm{sat}$ as a function of $l$ and $V_D$, respectively. The effective mobility ($\mu_\textrm{eff}$) shows a non-constant range profile for all the curves, since it depends on the accumulation layer thickness $l$. The mobility approaches to $\mu_\textrm{sat}$ after reaching the condition $l=D$. The range where $\mu_\textrm{eff}$ profile increases is attributed to the percolation regime where mobility increases with the rise of percolation paths \cite{ART_koehler2010}. When $\mu_\textrm{eff}$ reaches its saturation, it means that the accumulation layer is a constant ($l=D$) and the traps do not present more dependence on the transport. Most of the region where the $\mu_\textrm{eff}$ is non-constant, it is equivalent to the linear field effect mobility ($\mu_{\textrm{lin}}$).
    \par The output curves in Figure \ref{fig:EGOFET_2D} g) have two interesting results to be highlighted: (i), influence of percolation transport and (ii) non-linear behavior of the output curve for lower $V_D$ range. Considering (i), the influence of percolation transport: when the $\alpha$ parameter is varied, it is observed that the highest current intensity occurs for lower $\alpha$, which means it is observed a higher current for 2D PT. It occurs since the 2D PT has lower degrees of freedom than the 3D PT regime. The transport in 3D can occurs along the plane film and also inter-plane, increasing the path length taken by the charge carriers along the channel. As a consequence, the output current also shows lower intensity. The same explanation also justifies the difference among the intensities of the effective mobility in the graphs c) and e) in Figure \ref{fig:EGOFET_2D}. Considering (ii), even though the present theoretical development was based on ohmic contacts \cite{ART_koehler2010}, a non-linear behavior was observed in the output curve for lower $V_D$ range. This profile is frequently attributed to non-ohmic contacts  \cite{ART_waldrip2020,ART_liu2015}. In the present model, such behavior appears for the range of $\alpha \geq \gamma-1$  which corresponds to a situation more close to a narrow Gaussian distribution of localized states. Under this situation the Gaussian tail has a small amount of localized states that act like shallow traps states. More than this, close to the source the accumulation thickness ($l$) is quite thin (Figure \ref{fig:EGOFET_2D} a)). Looking back to Figure \ref{fig:EGOFET_2D} e), for the equivalent $V_D$ range analysis the $\mu_{\textrm{eff}}$ is quite low, which also reduces the output current intensity. Therefore, the output curve parabolic profile can be attributed to the injection limited current due to the very thin accumulation thickness close to the source along with a relevant amount of traps right after the injection interface. Such physical explanation to the non-ohmic output curve behaviour at low $V_D$ from a theoretical point of view is a novelty.
    \par The right column graphs in Figure \ref{fig:EGOFET_2D} depict an analogous simulation to the left one, but now for $\gamma=2.9$, represented by blue curves. This higher $\gamma$'s value simulate a higher amount of traps into the semiconductor in comparison to the previous analysis. The accumulation thickness as a function of $V_D$ in the figure \ref{fig:EGOFET_2D} b) does not depict relevant changes in comparison to figure \ref{fig:EGOFET_2D} a). In the figure \ref{fig:EGOFET_2D} d) and f), the effective mobility normalized to the $\mu_{\textrm{sat}}$ reaches values higher than one. It means that the linear field effect mobility ($\mu_{\textrm{lin}}$) is higher than the saturation one ($\mu_{\textrm{sat}}$). Figures \ref{fig:EGOFET_2D} d) and f) can be analyzed in the view of three regimes: (i) Percolation regime, where the mobility increases due to the rise of percolation paths; (ii) Bulk transport, where the mobility intensity decreases with the effective thickness rising along many monolayers \cite{ART_koehler2010}, resulting in a lower net electric field intensity leading to a lower average charge carriers velocity; (iii) Still bulk transport, but now with a constant $\mu_\textrm{eff}$ that occurs when $l=D$ along with the fact that there is no longer any influence of the traps. Figure \ref{fig:EGOFET_2D} h) shows the typical behavior of an output curve with a linear regime followed by saturation. 
    \par
\begin{figure}
        \centering
        \includegraphics[width=\columnwidth]{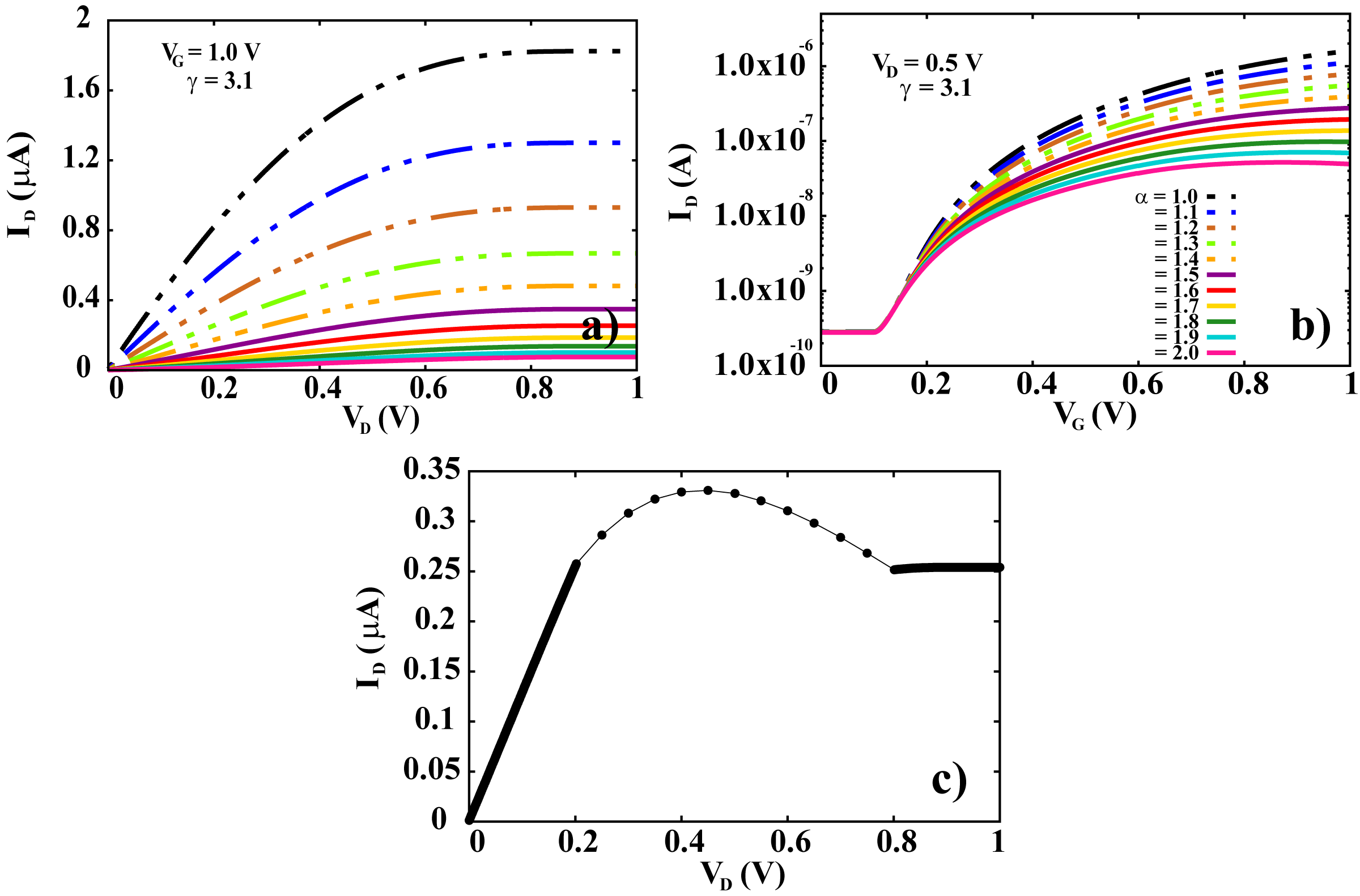}
        \caption{a) Output and b) transfer curves for a percolation transport in 2D (dashed lines) and a percolation transport in 3D (solid lines), where $\gamma = {3.1}$. The current intensity for $\alpha$ values equivalent to 2D transport is higher than for 3D case. c) Hypothetical transition from $\alpha = {1.3}$ to $\alpha = {1.6}$, where $\alpha$ has a step of {0.025}{} every {0.05}\,\rm{V} in $V_D$. Here $\gamma ={3.1}$, $V_D = 1.0\,V$ and $V_T = 0.1\,V$.}
        \label{transition_transport_2Dto3D}
    \end{figure}

 Interestingly, the current intensity depends on the kind of percolation transport: 2D or 3D, modeled when $\alpha$ parameter is changed. Figure \ref{transition_transport_2Dto3D} a) and b) shows the output and transfer curves, respectively, for different $\alpha$'s but keeping all other parameters constant. In the output curve, the typical linear and saturation regimes are observed and the current intensity decreases for increasing $\alpha$ values. The dash double-dot lines depict output curves under 2D PT range and the solid-lines for 3D one. In the transfer curve ( Figure \ref{transition_transport_2Dto3D} b)), relevant profiles changes are observed. While for $\alpha=1$ the current is still modulating when $V_G=1\,V$, providing us the information that a higher $V_G$ will still increase the on-off ratio e.g., for $\alpha=1.6$ the maximum modulation was already reached for $V_G=1\,V$ with a lower current intensity.
    \par As showed before, as higher is the voltage $V_D$, thicker will be the accumulation layer thickness (see graphs from Figure \ref{fig:EGOFET_2D} a) and b)). It means that a transition among growing values of $\alpha$ can occurs under a $V_D$ sweep, whose transient depends on the morphology of the film, not being possible to have a universal model for predicting this transition. The graph from Figure \ref{transition_transport_2Dto3D} c) depicts such situation where: $\alpha=1.3$ from $V_D=0.0\,V$ up to $0.2\,V$ and, $\alpha=1.6$ from $V_D=0.8\,V$ up to $1.0\,V$. We supposed to have an $\alpha$ transient from $1.3$ to $1.6$ with step of 0.025 for every 0.05 V in $V_D$, for the drain voltage range of $0.2$ up to $0.8\,V$. In this way, we manually extract each point that compose this region of the curve to illustrate such a situation. Due to the chosen $\alpha$ values, it represents a transition from 2D to 3D PT where a protuberance/lump is observed between the linear to the saturation typical regimes. Such profile has already been reported in experimental works but it has not yet been described using a 2D to 3D PT transition model. Such lump profile can also appear for $\alpha$ variation among just 2D PT range ($\alpha$=1 to 1.4) or just 3D one ($\alpha$=1.5 to 2).It is important to note that, from the set of parameters provided by the fit based on the present model, a microscopic analysis is made to understand some typically observed experimental behaviours.
\section{Experimental Fit}
\label{Exp_fit}
    \par
\begin{figure}
        \centering
        \includegraphics[width=\columnwidth]{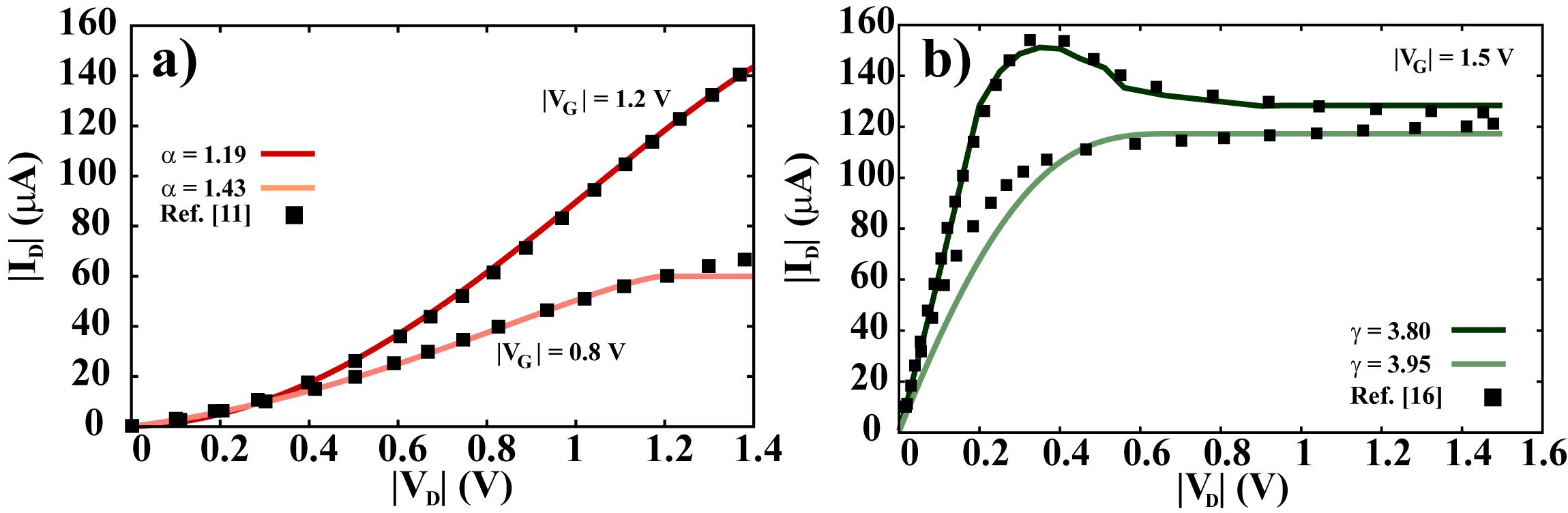}
        \caption{a) Experimental data (solid square dots) and theoretical fitting (solid lines) from an EGOFET output curves using the experimental parameters \cite{ART_kergoat2012_PNAS_non_linear_fit}: $|V_G|=1.2\,\rm{V}$ and $|V_G|=0.8\,\rm{V}$. The theoretical variable parameter: $\alpha =1.19$ for dark red and $\alpha =1.43$ for light red. b) Theoretical fit on experimental data from an OECT output curve \cite{ART_laiho2011}, considering a transient from $\alpha=1.375$ to $\alpha=1.600$, for $|V_G| = 1.5\,\rm{V}$, $|V_T| = 0.5\,V$ and $\gamma = 3.8$ (dark green) to forward part of the cyclic measurement. For the backward cycle (light green), it was used $\alpha = 1.6$, $|V_T| = 0.8\,V$, $\gamma = 3.95$.}
        \label{fig:experimental_fit}
    \end{figure}

 In the literature it is possible to find some papers reporting non-ohmic shaped output curves despite ohmic contacts. Figure \ref{fig:experimental_fit} a) shows our model theoretical simulation fit (solid line) performed on the output curves from the experimental data of \cite{ART_kergoat2012_PNAS_non_linear_fit} (solid square dots). The device is a poly-electrolyte-gated OFET depicting a non-linear behavior for low $V_D$ range, whose structure is: Ti-Au/P3HT/P(VPA-AA)/Au and the used parameters on the modelling were extracted from the experimental measurements \cite{ART_kergoat2012_PNAS_non_linear_fit}:  $V_T = {0.4}\,\rm{V}$ and $C_i = {1.0}\,\rm{\mu F/cm^2}$, while the channel dimensions are $W = 15\,\rm{mm}$, $L = 2\,\rm{\mu m}$ and $D = 30\,\rm{nm}$. The theoretical parameters are: $\alpha = {1.19}{}$ (for $|V_G| = {1.2}\,\rm{V}$, dark red line) and $\alpha = {1.43}{}$ (for $|V_G| = {0.8}\,\rm{V}$, light red line), $\kappa = {4.0}{}$, $T = {300}\,\rm{K}$, $D_c = {0.4}\,\rm{nm}$, $\mu_\textrm{sat} = {0.2}\,\rm{cm^2/Vs}$ and $\gamma = {1.8}{}$. Since $\alpha$ is related to a depth of percolation path, it seems coherent that this parameter changes depending on $V_G$, once a higher gate voltage produces a higher electric field intensity and a thinner accumulation layer thickness at the interface. Therefore, we assume a lower $\alpha$ for higher $V_G$ (and \textit{vice-versa}) that results in a good fitting for both curves. For the light red line fit, there is no agreement between experimental and theoretical data for $V_D>|1.2|\,\rm{V}$. From our simulation, this curve must saturate after $\sim 1.2\,\rm{V}$, since it was used the extracted experimental $V_T=+0.4\,\rm{V}$ and therefore we do not interpret it as an inappropriate adjustment. From this fit, microscopic parameters were obtained: $\gamma$ which indicates a narrower Gaussian distribution of states and $\alpha$ which is attributed to 2D percolation transport profile.
    \par The experimental data in figure \ref{fig:experimental_fit} b) is from a polyelectrolyte-gated organic thin-film transistors with structure: Au/P3CPT/P(VPA-AA)/Ti \cite{ART_laiho2011} with a clockwise cyclic curve. The theoretical fit was separately simulated for the forward measure (dark green solid line) and backward one (light green solid line). The authors of the experimental work analyzed the transistor mode of operation considering ions diffusion into the channel, that is, it operates as an organic electrochemical transistor (OECT). The experimental data \cite{ART_laiho2011} and theoretical parameters used were: $W = {1}\,{mm}$, $L = {3}\,{\mu m}$ and $D = {30}\,\rm{nm}$, $C_i = {4}\,{\mu F/cm^2}$, $|V_G| = {1.5}\,\rm{V}$  and $|V_T| = {0.5}\,\rm{V}$, $\kappa = 4.5$, $T ={300}\,\rm{K}$, $D_c ={0.3}\,\rm{nm}$, $\mu_\textrm{sat} = {6.5\times 10^{-3}}\rm{cm^2/Vs}$ and $\gamma = {3.8}{}$. To fit the protuberance profile between the linear and saturation regimes in the forward measure, it was considered a transient from $\alpha = 1.375$ up to $\alpha = 1.600$ in the region where the lump profile appears, resulting in a very good fit.
    \par Despite the present model is fully based on the description of transistors based on field effect modulation like FETs, OFETs or EGOFETs, we successfully fitted an OECT describing a non-ideal behavior where a lump appears between the linear and saturation regimes. The interpretation given to this fit is that when ions diffuse into the channel resulting in ionic doping, they generate a new electrical rearrangement which can be represented by an equivalent local field effect. This distribution of charges and ions is found along the accumulation layer that can extend close to the surface of the semiconductor or along its entire volume. Similar situation, of non-constant hole and cation concentrations along the channel, has already been described in the literature where the cation concentration increases exponentially towards the drain electrode \cite{OECT_channel_thickness_exponential}. However, our model quantifies this field effect generated by both charge carriers and ions, without making a distinction between each one and providing a net result of their contributions. Therefore, while some OECTs models propose to quantify a volumetric capacitance generated by the diffusion of ions inside the channel, the present model shows that a simplified analysis is enough to quantify the field effect generated by both charges and ions along the accumulation layer. 
    \par It is important to note that after obtaining the simulation parameters from the output curve fitting, it is possible to infer informations not provided by the experimental characterization by plotting the graphs: $l-V_D$; $\mu_{\textrm{eff}}/ \mu_{\textrm{sat}} - V_D$; and electric field intensity as a function of $V_D$ or $y$, where $y$ represents the channel thickness direction. The mentioned graphs analysis are depicted in the supplementary materials using the parameters from the OECT fit simulation, figure \ref{fig:experimental_fit} b). It shows that the accumulation layer reaches the semiconductor thickness at $V_D\sim 1\,\rm{V}$. The $\mu_{\textrm{eff}}$ as a function of $V_D$ has three regimes: increasing, decreasing and saturation, whose description of the physical phenomena are the same as those attributed to the figure \ref{fig:EGOFET_2D} f). According to our model, in the range of the linear regime $\mu_{\textrm{lin}}$ it has a maximum value at $V_D\sim 0.4 V$ and $\alpha=1.375$, almost hundred times higher than $\mu_{\textrm{sat}}$, and at $V_D\sim 0.8 V$ with $\alpha=1.600$ it is almost twenty times higher than $\mu_{\textrm{sat}}$. In a parallel analysis of the mobility intensity on the graph of the electric field intensity along the accumulation layer, it is possible to perceive a region with an electric field a hundred times higher in a region that goes from the source to part of the channel length, reducing closer to the drain. This region with such higher electric field intensity is attributed to the region with the highest concentration of ions diffused into the channel. In summary, this high concentration of charges and ions when there is a very thin accumulation layer results in such a very high field effect mobility. The in-homogeneity of ions distribution along the channel has already been shown through a non-quantitative measurement named as simple $I - V$ curve in the literature \cite{Seidel_2021_transistor_Elton}, where the curve profile is asymmetric showing that ions are not equally distributed within the channel under certain conditions. 
    \par For the backward cyclic measure direction (from figure \ref{fig:experimental_fit} b), it was necessary to increase the values of two parameter: $\gamma=3.95$ and $V_T=0.8\,\rm{V}$. Since ions have been diffused during the measurement performance it is expected that the properties of the semiconductor/electrolyte interface are changed. In this case, the new parameters provide the information that there are more traps close do the interface and a broader Gaussian distribution that is coherent with the presence of ions into the channel.
    \par Finally, This model proved capable of bringing physical interpretations of the operating modes of EGOFETs and OECTs. In addition to extracting typical transistor parameters, microscopic information is provided from the theoretical adjustment, inferring information not obtained only through electrical measurements.
\section{Conclusion}
    \par A theoretical model was proposed for the charge carriers transport in electrolyte-gated transistors (EGTs) considering the influence of a shallow exponential traps distribution in the semiconductor and 2D or 3D percolation transport (PT). An important inference of the model is that the influence of the high capacitance per unit area formed at the interface semiconductor/electrolyte dielectric gate promotes changes on the thickness of the accumulation layer. The accumulation thickness does not necessarily extend along the entire semiconductor thickness. This influence was inserted into the model in the effective mobility parameter ($\mu_{\textrm{eff}}(l)$). It is possible to analyze both EGOFETs and OECTs modes of operation in steady-state and extract parameters as: field effect mobility; $\gamma$ that correlates to the exponential energetic depth of traps; and $\alpha$ that provides if 2D or 3D PT occurs. These parameters were used to simulate: (i) accumulation layer thickness profile and (ii) effective mobility, both as function of $V_D$. The last provides an interpretation if the linear field effect mobility $\mu_\textrm{lin}$ has a non-constant behavior and if it is lower or higher than $\mu_\textrm{sat}$. For an EGOFET, the accumulation layer thickness describes the charge carriers distribution along the channel length, while for OECTs it represents the distribution of both charge carriers and diffused ions. The present model considers just the net electric field formed by the distribution of both charges carriers and ions, without distinguishing their contributions. 
    \par Two theoretical/experimental fits were demonstrated: (i) the non-linear behavior for low $V_D$ in an EGOFET output curve considering an ohmic contact. Such profile is attributed to the very thin effective thickness ($l$) close to the source together with the influence of the exponential traps presence and this interpretation extracted from a model simulation is a novelty; (ii) the adjustment of the lump behavior between the linear and saturation regimes observed on some EGTs output curves. The lump profile was attributed to a transient from 2D to 3D PT under the analyzed parameters that were reasonable estimates. Based on these two successful adjustments, the proposed model can be considered general to be applied in EGTs operating as EGOFET or OECT.
    \begin{acknowledgments}
        The authors thank to the Coordenação de Aperfeiçoamento de Pessoal de Nível Superior – Brasil (CAPES) – Finance Code 001.
    \end{acknowledgments}


\bibliography{apssamp}

\end{document}


\preprint{APS/123-QED}
\title{Supplementary Material\\
General model for charge carriers transport in electrolyte-gated transistors}

\author{Marcos Luginieski}
 \affiliation{Instituto de Física de São Carlos, Universidade de São Paulo, CP 369, CEP 13660-970, São Carlos, SP, Brazil}
  \affiliation{Universidade Tecnológica Federal do Paraná - UTFPR, Av. Sete de Setembro, 3165, CEP 80230-901 Curitiba, Brazil}
 \email{mluginieski@gmail.com }
\author{Marlus Koehler}%
 \affiliation{Universidade Federal do Paraná - UFPR, Centro Politécnico, Jardim das Américas CP 19044, CEP 81531-990 Curitiba,
Brazil}
%
\author{José P. M. Serbena}
 \affiliation{Universidade Federal do Paraná - UFPR, Centro Politécnico, Jardim das Américas CP 19044, CEP 81531-990 Curitiba,
Brazil}%
\author{Keli F. Seidel}
\email{keliseidel@utfpr.edu.br}
 \affiliation{Universidade Tecnológica Federal do Paraná - UTFPR, Av. Sete de Setembro, 3165, CEP 80230-901 Curitiba, Brazil}%


\date{\today}

\maketitle
\section{Traditional Organic Field Effect Transistors' Equations}
\par
\begin{figure}[h!]
    \centering
    \includegraphics[width=1\linewidth]{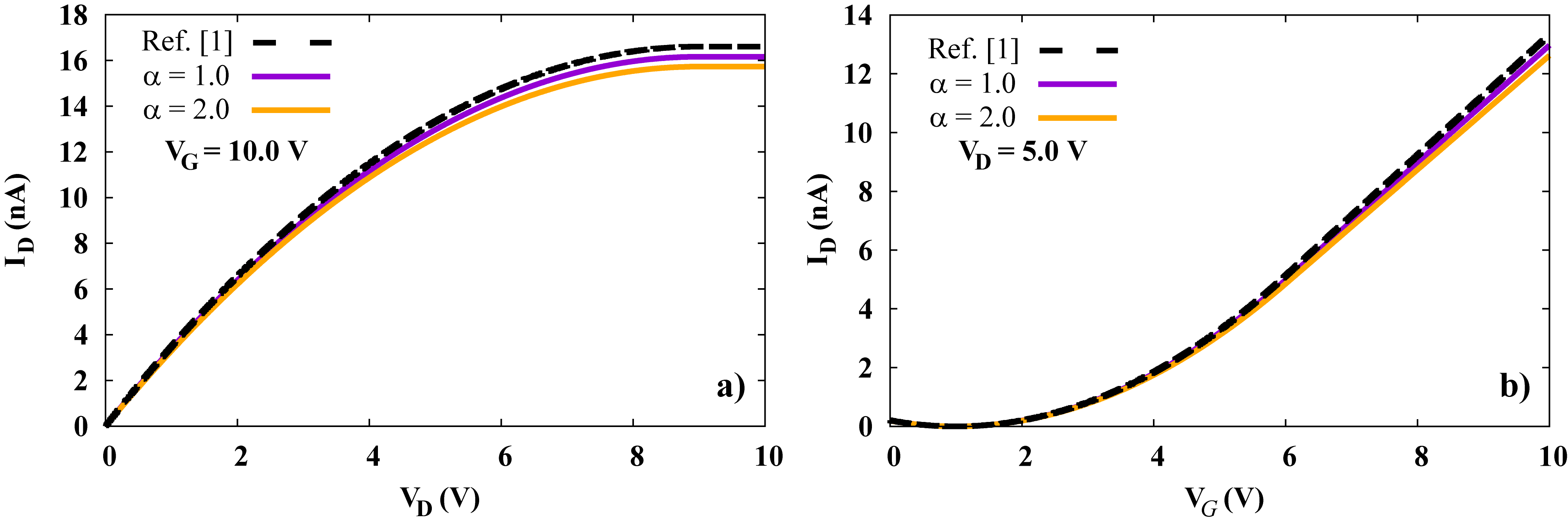}
    \caption{OFET's output a) and transfer b) curves for equations \eqref{Eq:ID_nosso_OFET}, \eqref{Eq:Isat_nosso_OFET} and those from ref. \cite{inbook_horowitz2010}. The parameters used are the same as those from figure 2 in the main text. The exception are $V_T = 0.2\,\rm{V}$ and the $\alpha$ depicted in the graphs.}
    \label{fig:ID_OFET}
\end{figure}

 For typical OFET capacitances ($\sim\,nF/cm^2$) and for a thickness $D$ of dozens of $nm$, $V_\textrm{tr}$ can be larger than $V_\textrm{sat}$, then $V'' < 0$. In this case, every positive bias satisfies the condition from equation (2) of main text, where $l = D$. With that, the effective field effect mobility is given only by equation (4) of main text. Plugging this equation into the (5), one has
\begin{align}
    I_{D} = \frac{W C_i}{L}\int_{0}^{V_{D}} \mu_\textrm{sat}\left(\frac{D - D_{c}}{D}\right)^{\alpha} [V_G - V_{T} - V(x)] dV,
\end{align}
or
\begin{align}
    I_{D} = \frac{W C_i}{L}\mu_\textrm{sat}\left(\frac{D - D_{c}}{D}\right)^{\alpha} \left[(V_G - V_{T})V_D - \frac{V_D}{2}\right]V_D.
    \label{Eq:ID_nosso_OFET}
\end{align}
Being $V_\textrm{sat} = V_G - V_T$, for the saturation regime the last equation becomes
\begin{align}
    I_\textrm{sat} = \frac{W C_i}{L}\mu_\textrm{sat}\left(\frac{D - D_{c}}{D}\right)^{\alpha} (V_G - V_{T})^2.
    \label{Eq:Isat_nosso_OFET}
\end{align}
Both of the latter equations resemble the traditional OFET equations \cite{inbook_horowitz2010}, in that the only difference is the $[(D-Dc)/D]^\alpha$ term. For typical values of $D_c$ (a few nanometers), it occurs that $D \gg D_c$, so that this term tends to 1. Thus, in this limit the equations above become exactly the same as those originally derived for OFETs \cite{inbook_horowitz2010}. In figure \ref{fig:ID_OFET} it is possible to see output and transfer curves simulated with Eqs. \eqref{Eq:ID_nosso_OFET} and \eqref{Eq:Isat_nosso_OFET}, in comparison with the traditional OFET model ones.
\section{Experimental Fit}
\par
\begin{figure}[h!]
        \centering  \includegraphics[width=\columnwidth]{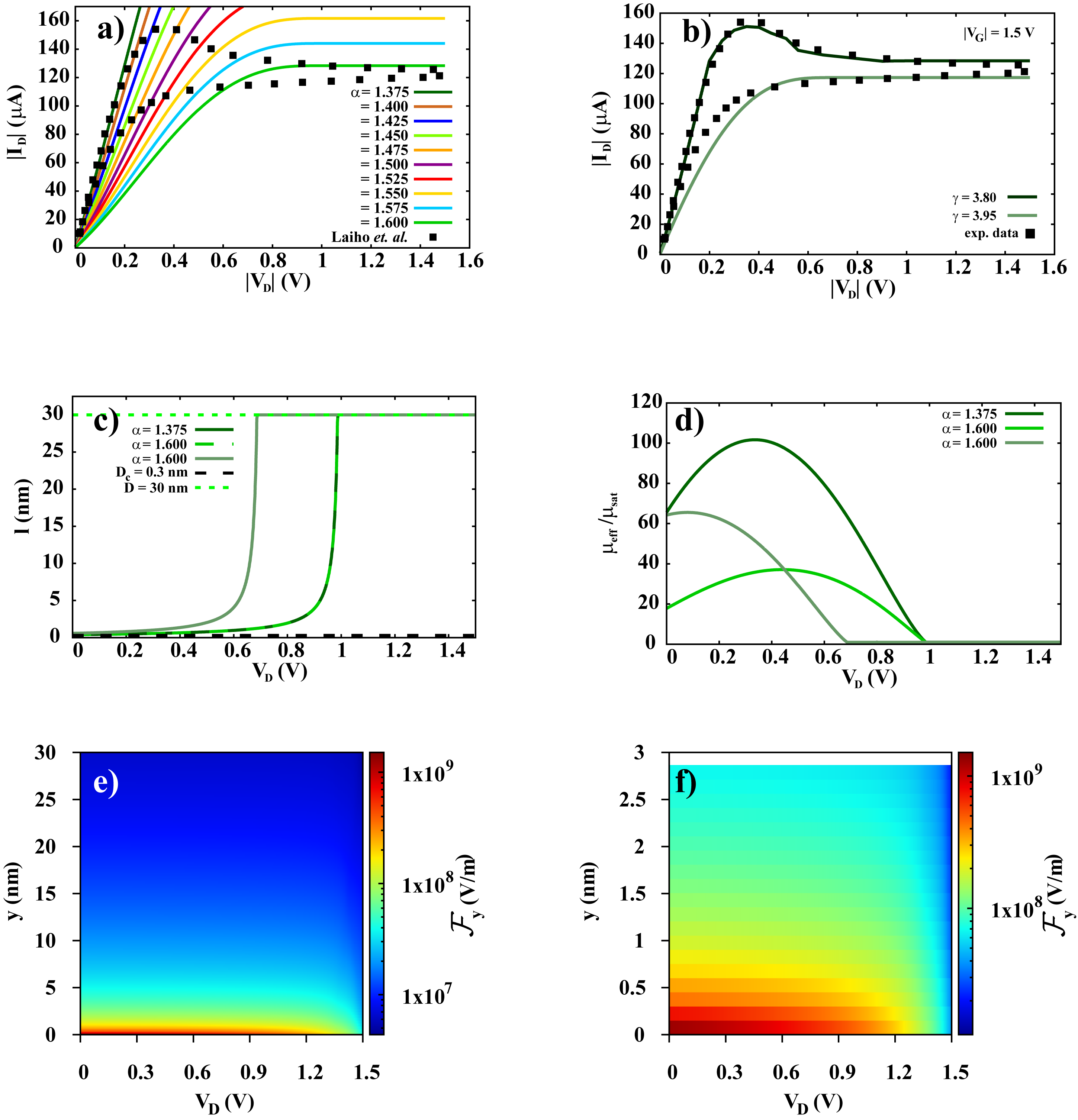}
          \caption{a) and b) Theoretical fit on experimental data from an OECT output curve, considering a transient from $\alpha=1.375$ to $\alpha=1.600$, for $|V_G| = 1.5\,V$, $|V_T| = 0.5\,V$ and $\gamma = 3.8$ (dark green) to forward part of the cyclic measurement. For the backward cycle (light green), it was used $\alpha = 1.6$, $|V_T| = 0.8\,V$, $\gamma = 3.95$. In the sequence, simulation graphs based on the fitted parameters from curve (b): c) $l-V_D$, d) ($\mu_\textrm{eff}/\mu_\textrm{sat}) - V_D$, e) Channel thickness ($y$) and Electric field profile along the channel and f) Zoom from graph e). }
        \label{fit2}
    \end{figure}

 Experimental data (square dots) in figure \ref{fit2} a) and b) is from a polyelectrolyte-gated organic thin-film transistors with structure: Au/P3CPT/P(VPA-AA)/Ti \cite{ART_laiho2011} with a clockwise cyclic curve. After obtaining the simulation parameters from the output curve fitting (from figure \ref{fit2} a) and b)), we can predict information not provided by the experimental electrical characterization, plotting the graphs as: $l-V_D$ (figure \ref{fit2} c)); $\mu_{\textrm{eff}}/ \mu_{\textrm{sat}} - V_D$ (figure \ref{fit2} d)); and the electric field intensity as a function of $V_D$ or $y$ (figure \ref{fit2} e) and its zoom in figure \ref{fit2} f)), where $y$ represents the thickness channel direction. 

%

%
%
\newpage
\bibliography{apssamp}